\documentclass[12pt,preprint]{aastex}

\shorttitle{Hercules Thick Disk Cloud}
\shortauthors{Larsen et al.}

\begin{document}

\title{Mapping the Asymmetric Thick Disk:  The Hercules Thick Disk Cloud}

\author{Jeffrey A. Larsen}
\affil{Physics Department, United States Naval Academy,
    Annapolis, MD 21402}
\email{larsen@usna.edu}

\author{Roberta M. Humphreys}
\affil{Astronomy Department, University of Minnesota, Minneapolis MN, 55455}
\email{roberta@umn.edu}

\and

\author{Juan E. Cabanela}
\affil{Department of Physics and Astronomy, Minnesota State University Moorhead, Moorhead MN, 56563}
\email{cabanela@mnstate.edu}

\begin{abstract}
The stellar asymmetry of faint thick disk/inner halo stars in the first 
quadrant ({\it l} = 20 --45$\arcdeg$) first
reported by \cite{lar96} and investigated further by  \cite{par03,par04} has 
recently been
confirmed by the SDSS \citep{jur08}. Their interpretation of the 
excess in the star counts as a 
ringlike structure, however, is not supported by 
critical complementary data in the fourth quadrant, 
not covered by the SDSS.  We present stellar 
density maps from the Minnesota Automated Plate 
Scanner (MAPS) Catalog of the POSS I showing that 
the overdensity does not extend into the fourth 
quadrant. The overdensity is most probably not a ring. It
could be due to interaction with the disk bar, 
evidence for a triaxial thick disk, 
or a merger remnant/stream. 
We call this feature the Hercules Thick Disk Cloud.

\end{abstract}

\keywords{Galaxy: structure, Galaxy: kinematics and dynamics}

\section{Introduction}

\cite{lar96} initially reported a substantial asymmetry of faint blue stars 
in the first quadrant (Q1)
of the inner Galaxy,  $l = 20\arcdeg - 45\arcdeg$ compared with 
complementary fields in the fourth 
quadrant (Q4) based on star counts from MAPS\footnote{The Minnesota Automated Plate 
Scanner Catalog of the POSS I is online 
at:  http://aps.umn.edu} \citep{cab03}.
\cite{par03} made a more in-depth survey to map the extent 
of the asymmetry using
40 contiguous fields in each of three regions:  Q1 above and 
below the plane and Q4 above the plane.   Q4 below the plane 
is not covered in the POSS I.  They found a
25\% excess in the number of  probable thick disk stars in Q1 
above \it{and }\rm below the plane when compared to the 
complementary Q4 fields.  The region was irregular in shape 
and covered several hundred square degrees, but with a completeness
limit at $\approx$ 18--18.5  mag, the stars showing the excess in Q1 were
relatively nearby, $\sim$ 1 -- 2 kpc from the Sun.  \cite{par04} also found an
associated kinematic signature, a significant lag  of  ~80 to 90 km/sec in the
direction of Galactic rotation for the associated thick disk stars in Q1.

The recent release of the SDSS Data Release 5 (DR5) photometry in
the direction of the observed asymmetry in
Q1 led to the discovery of a feature at much fainter magnitudes, 
the distant Hercules-Aquila
cloud \citep{bel07} 
and the photometric parallax study by \cite{jur08} confirmed our nearer asymmetry
in the inner Galaxy as an overdensity at a galactocentric radius of 6.5 kpc situated
1.5 kpc above the plane.  
The SDSS survey however is not
well designed for a good study of the thick disk inside the Solar orbit. It
extends below $b = 30\arcdeg$ in only a few directions in Q1 and has only 
limited coverage in Q4.

We are continuing our program of photometric and spectroscopic 
observations to map the size and extent of the asymmetry along our line of sight
and to determine the degree of spatial and kinematic asymmetry above and below 
the plane.  We are using the SMARTS Consortium CTIO 1-meter Y4KCam and the 
Steward Observatory 90" Bok telescope 90Prime Mosaic imager to obtain 
wide-field multi-color CCD imaging to fainter completeness limits than 
the POSS I.  Spectra of the candidate thick disk stars for radial velocities 
and metallicity estimates have
been observed using the Hydra multi-object spectrometer on the  
CTIO Blanco 4-meter telescopes and the Hectospec on the MMTO 6.5 meter.

In this Letter we present a stellar number 
density map we created from the MAPS POSS I scans to provide a more global
reference for our current deeper but more spatially restricted photometric and 
spectroscopic study of the thick disk in the inner Galaxy.  Other 
works \citep{xu07} have used plate data to supplement gaps in the SDSS 
at more southern declinations with good success. 
Our map of the stellar density in Q1 and Q4  covers much of the sky 
unavailable to SDSS and  demonstrates that the nearby asymmetry in Q1 does 
not represent a ring above the Galactic plane, 
but instead is a significant substructure or cloud extending over many 
square degrees in galactic longitude and latitude.
We call this feature the Hercules Thick Disk cloud.

\section{The Stellar Density Maps}

\begin{figure*}
  \begin{center}
    \leavevmode
	\epsscale{0.6}
	\plotone{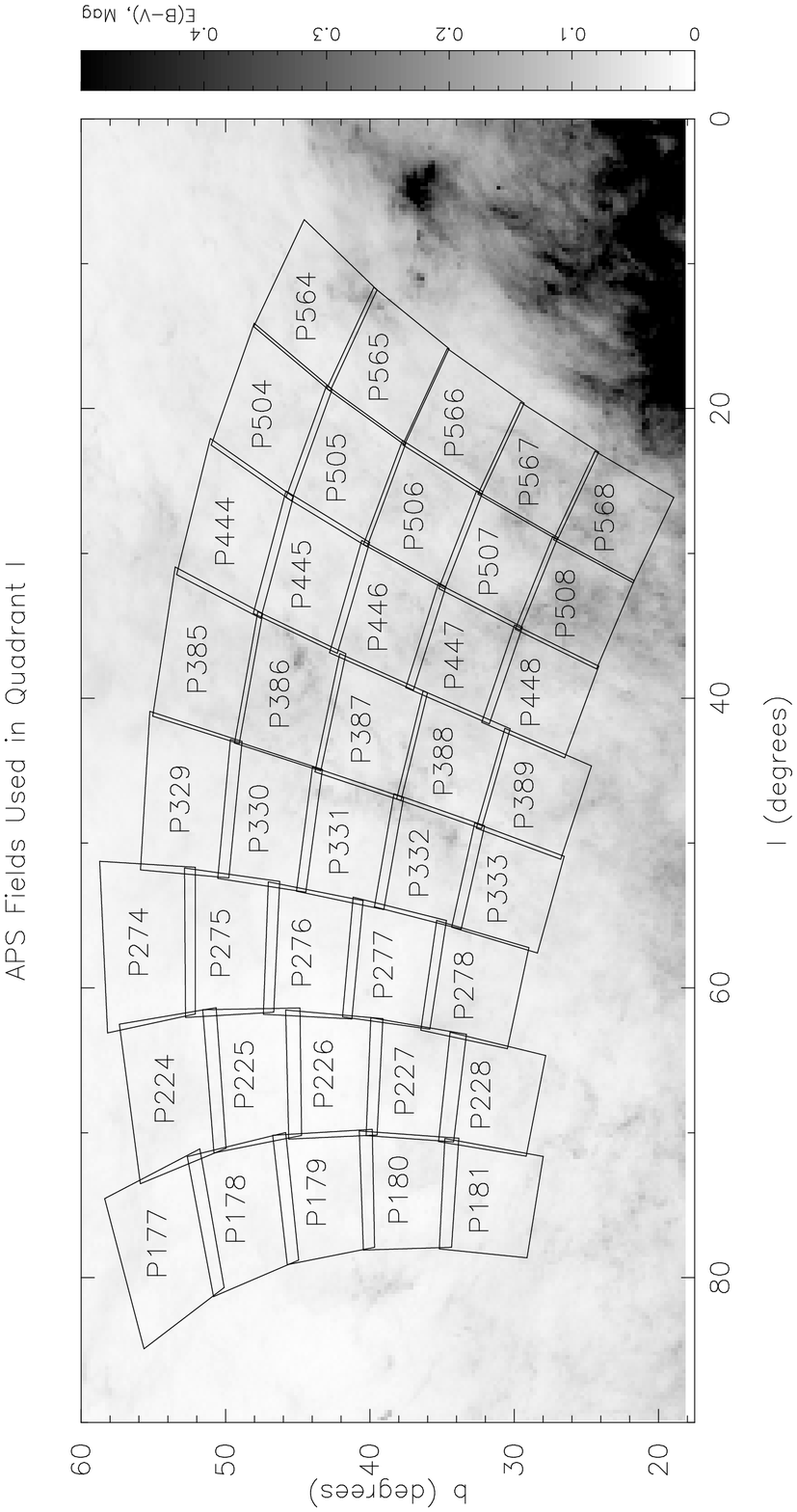}
      \caption{The fields from the MAPS Catalog of POSS I used to create the star count areal density image for Q1 ($l  = 0\arcdeg - 90\arcdeg$).  The centers are further described in \cite{par03} and are superimposed on the extinction plots of \cite{sch98}.  While the interstellar extinction is clumpy it is not high enough to influence our primary result.}
     \label{fig1}
  \end{center}
\end{figure*}

\begin{figure*}
  \begin{center}
    \leavevmode
	\epsscale{0.6}
	\plotone{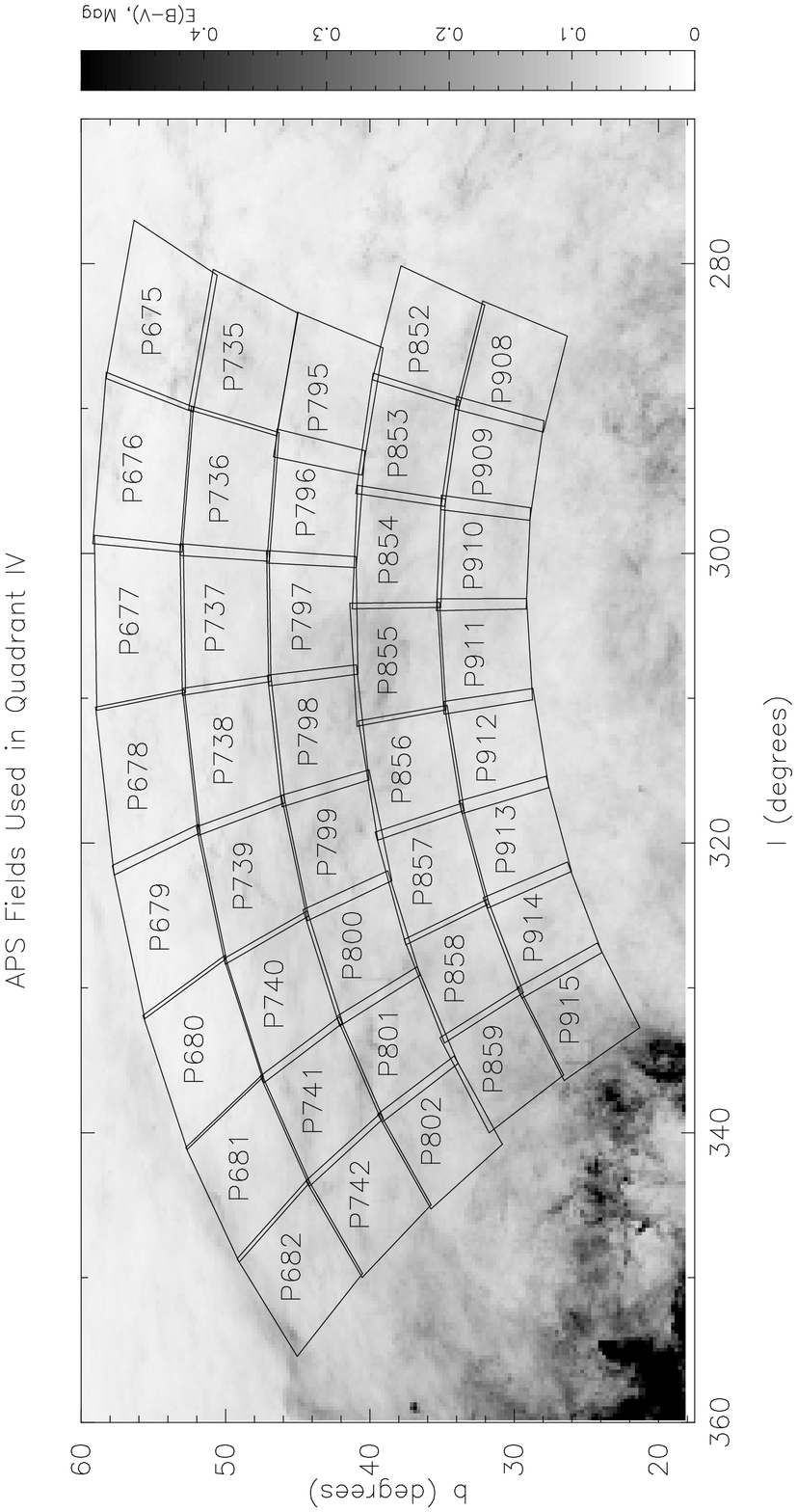}
      \caption{The fields from the MAPS Catalog of POSS I used to create the star count areal density image for Q4 ($l  = 270\arcdeg - 360\arcdeg$).  The centers are further described in \cite{par03} and are superimposed on the extinction plots of \cite{sch98}.  While the interstellar extinction is clumpy it is not high enough to influence our primary result.\label{fig2}}
  \end{center}
\end{figure*}

The figures presented in this Letter were originally created to provide a 
reference frame for the interpretation of our ongoing observations 
of the thick disk using narrower yet deep multicolor CCD imaging.  
They were made 
using the same set of POSS I fields selected by \cite{par03} above the 
galactic plane 
and the fields are fully described in that publication.   The 80 
POSS I plates cover 2900 square
degrees on the sky and all are complete to fainter than 18th magnitude in the
O band (blue) and have colors (O--E) available from the paired observation in
the red (E band). Their placement on the sky is shown in Figure~\ref{fig1} for
Q1 and for Figure~\ref{fig2} for Q4.  The B-V extinction in each field is
plotted from \cite{sch98} and is substantially less than $E(B-V) < 0.2$ in 
all fields.   We defer a discussion of the Q1 data below the plane to a 
later paper because it is not relevant to the present Letter. The MAPS 
Catalog has superior stellar photometry
when compared with most  other digitized plate  catalogs of the POSS 
because each
plate has its own independent photometric calibration derived from our own
CCD observations as described in \cite{lar03} or from photoelectric photometry
\citep{las88} and uses an  isophotal diameter-to-magnitude relation for the 
stars.

The individual plate catalogs were merged, duplicates in the plate 
overlap regions
removed, and interstellar extinction from \cite{sch98} was applied on 
a star by star
basis using the standard interstellar extinction law.  We assume
that the bulk of the extinction comes from relatively near the Sun 
and that our objects
of interest are much farther away, so that the extinction is
a zero point correction.  A global geometric vignetting correction 
was applied to all of the plates
in the MAPS Catalog. However, we found that some plates had larger 
vignetting problems (probably due to moonlight).  
As a result we applied an additional radial magnitude
correction to the MAPS magnitudes for 5 of the 80 plates to bring
the number densities at the plate edges into line with the well behaved and
well calibrated plate center.  Star-galaxy classification uses a neural 
network \citep{1992AJ....103..318O,1993PASP..105.1354O} and is described 
in those papers and in \cite{cab03}.  Any uncertainty in object 
classification is not
a significant factor in the creation of these maps.

\begin{figure*}
  \begin{center}
    \leavevmode
	\epsscale{0.7}
	\plotone{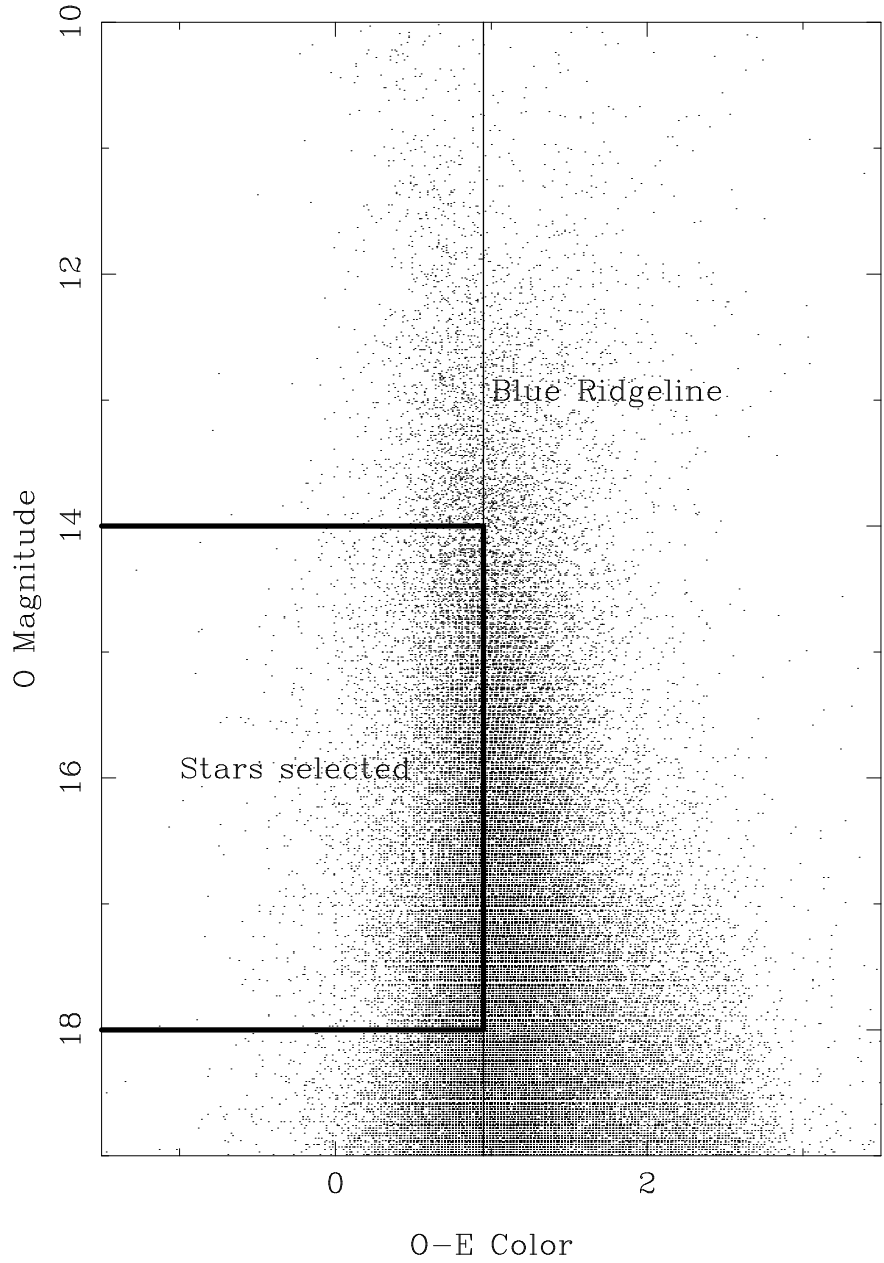}
\caption{An example Color Magnitude diagram from the MAPS Catalog indicating the region of color and magnitude used to generate these images.\label{fig4}}
  \end{center}
\end{figure*}

We then selected stars in the magnitude range $14 < O < 18$ (or
approximately $13.5 < V < 17.5$) above the completeness limits.
Since the plates still have some magnitude zero point differences and cosmic
scatter, we used the  ``blue ridgeline" at 
$O-E \approx 1.0$ or $B-V \approx 0.6$ as a fiducial reference 
to select a sample of stars with blue and intermediate 
colors \citep[see][Fig.5]{par03}.  Stars bluer than the ridgeline 
are representative of  the stars
which show the asymmetry \citep{par03}.   While the exact location
of the blue ridge in  O-E color varies somewhat from plate to plate and with
galactic longitude and latitude, due to the relative contributions of 
stars from the halo, disk and
thick disk, it provides 
a strongly identifiable feature on each color-magnitude
diagram (see Figure~\ref{fig4}).  The color variations between
adjacent plates will thus be small compared to the variations across the
full 180 degrees of sky covered.  Our magnitude and color search limits
are 
illustrated on one of our color magnitude diagrams shown in Figure ~\ref{fig4}.

\section{Discussion and Conclusions}

\begin{figure*}
  \begin{center}
    \leavevmode
	\epsscale{0.4}
	\plotone{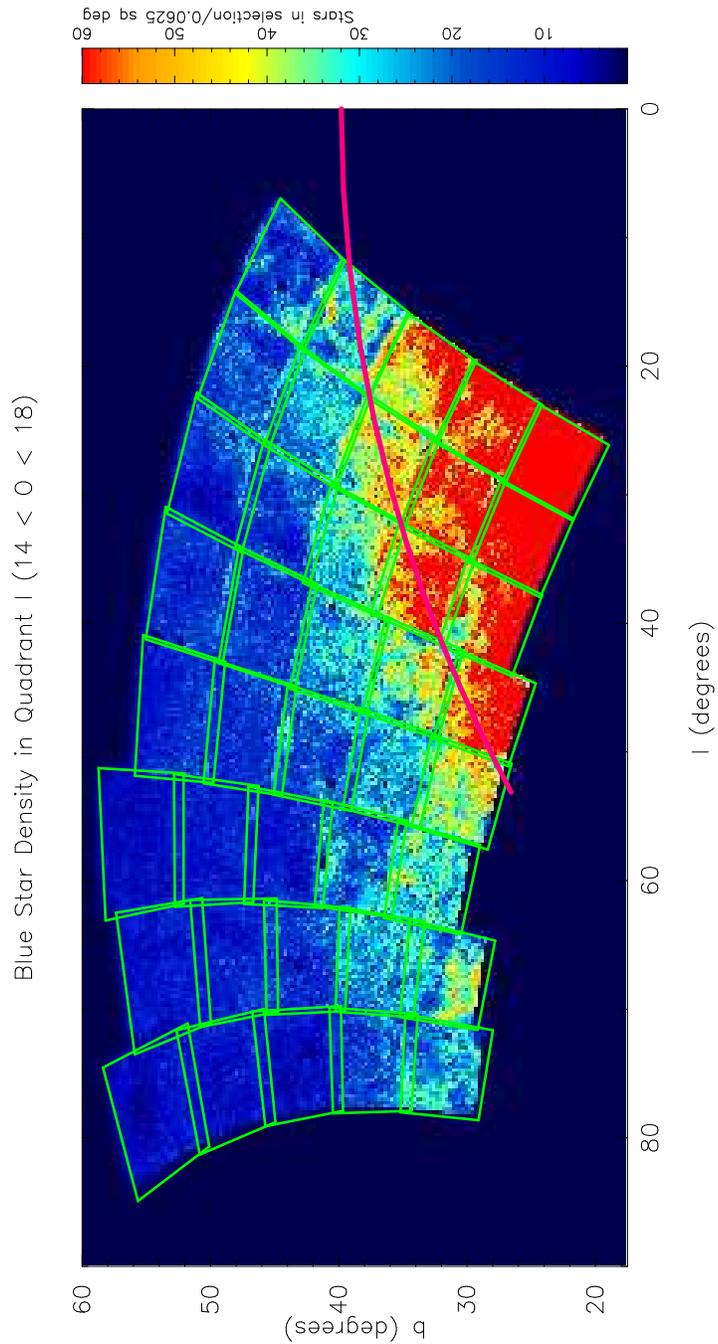}
      \caption{The density image created for the stars in Q1 with the POSS I plate boundaries from Figure ~\ref{fig1} overlayed in green.  Notice how there is a distinct overdensity on the order of about 30\% between $l=25\arcdeg - 45\arcdeg$ and $b = 30\arcdeg - 35\arcdeg$ when compared with the corresponding Q4 stars in Figure ~\ref{fig6}.  Comparison with Figure ~\ref{fig1} shows that extinction cannot be causing the feature we see.  The location of Juri{\c'} et al.'s overdensity is shown by the purple line.  \label{fig5}}
  \end{center}
\end{figure*}

\begin{figure*}
  \begin{center}
    \leavevmode
	\epsscale{0.4}
	\plotone{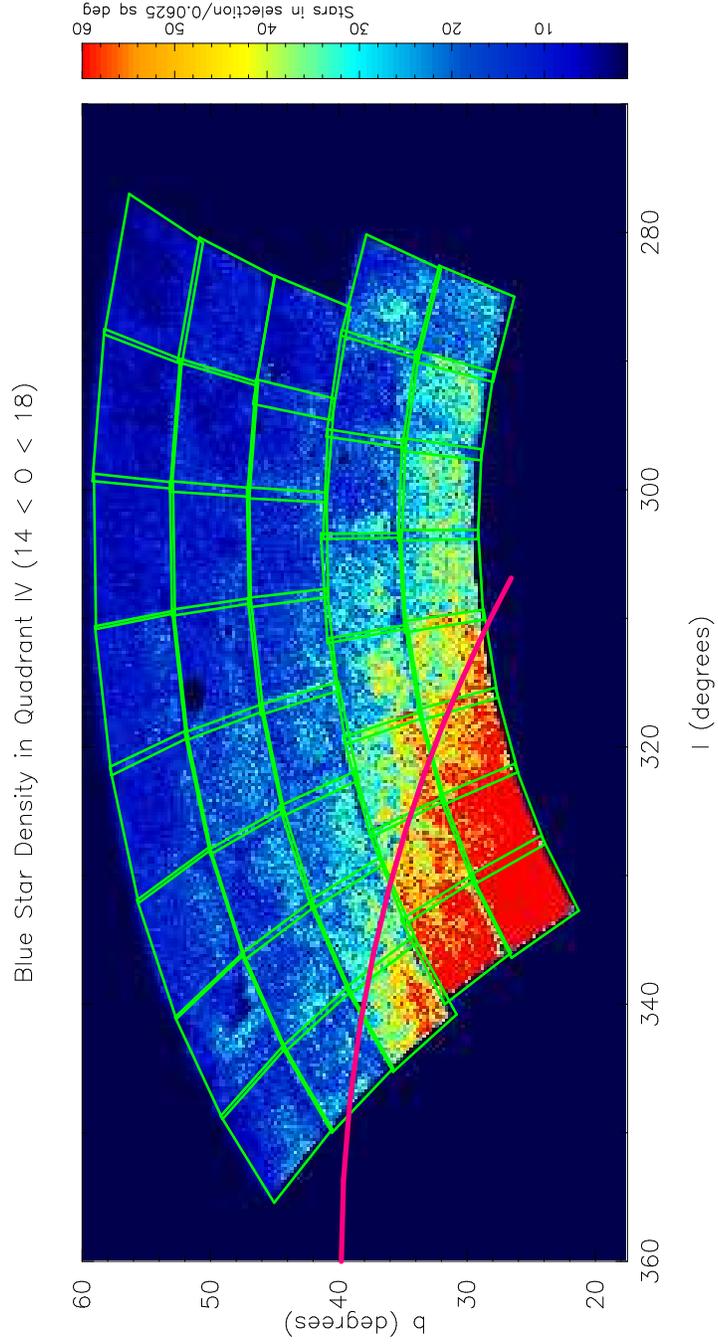}
      \caption{The density image created for the stars in Q4 with the POSS I plate boundaries from Figure ~\ref{fig1} overlayed in green.  There is no enhancement of faint blue star counts between  $l=335\arcdeg - 315\arcdeg$ to match the excess shown in Figure ~\ref{fig5}.  Comparison with Figure ~\ref{fig2} shows that extinction cannot be hiding this lack of stars.  The location of the Q4 projection of Juri{\c'} et al.'s overdensity is indicated by the purple line.  \label{fig6}}
  \end{center}
\end{figure*}

After the reduction steps outlined above, we then 
binned the stars $0.25 \arcdeg \times 0.25 \arcdeg$ 
in {\it l} and {\it b} to create maps of the stellar 
density distribution of the faint blue 
and intermediate color stars shown in
Figures~\ref{fig5} and ~\ref{fig6}. The figures are color-coded with respect to 
number density per 0.0625 square degree. 

\cite{jur08} described the excess 
in Q1 as due to a ``ringlike" 
structure because the overdensity 
appeared to be radially constant and circular in cross
section in their Figure 27.  
The center of the overdensity region
ranges from $(X,Y,Z)=(6.5 kpc, -2.2 kpc, 1.5 kpc)$ to 
$(6.5 kpc, 0.3 kpc, 1.5 kpc)$ and can easily be converted into galactic 
coordinates. This is shown as the purple line on Figure~\ref{fig5}.
If the feature were symmetric about $l=0\arcdeg$ it would project
into Q4 to $(6.5 kpc, 2.2 kpc, 1.5 kpc)$.  This projection is also shown
as a purple line on Figure~\ref{fig6}.
Comparison of Figures~\ref{fig5} and ~\ref{fig6}, however show that 
the density of these stars is not symmetric with 
respect to the Sun-Center line. There is a clear 
excess of stars in Q1 over Q4 in the range 
{\it l} $= 25$ -- $45\arcdeg$ and {\it b} $= 30$ -- $40\arcdeg$.
Furthermore, we emphasize that the excess would not have been initially 
discovered if
it had been a symmetric ring since we \citep{lar96} were 
comparing star counts for 
complementary fields in Q1 and Q4.

Could the ``ringlike" structure be inclined to the plane and therefore
not visible in Q4?  Most probably not.  The full height of the overdensity 
in Z over an azimuthal distance of 2500 parsecs is only 500 pc (Juric{\c'} et al.). 
Even for the pathological case of
a paper thin inclined distribution in Z the maximum inclination could only be 
11 degrees and given its width in X it should have been visible in Figure~\ref{fig6}.  
Additionally, there is strong evidence from Juri{\'c} 
et al.'s Figure 27 (left panel) that for X = 7250-7750 parsecs the
overdensity is exclusively above $Z = 1500$ pc.  Examination of the same figure's
middle panel shows that all significant contributors to the overdensity in 
this same X range have $Y < 1000$ pc.  This would be 
the opposite of what should be happening if the overdensity were falling into
the plane in Q4.  Finally, \cite{par04} studied the velocities of
samples of stars taken from overdensity regions in Q1 compared with a control 
sample Q4 and found in a somewhat weak result that the Z component of 
velocity was less negative for Q1 than it was for Q4.
If a coherent population of stars were moving together towards the disk from
above the plane as they entered Q4, the opposite should be true.

The cloud of stars detected by Juri{\'c} et al. is 
not symmetric about the $l=0\arcdeg$ line and almost certainly is
not a ring.
This does not change their other possible 
explanation for the feature, however.  Given the broad extent of the cloud 
(Figures~\ref{fig5} and ~\ref{fig6}) 
together with its apparent small ranges in radial distance
from the Sun and its distance above the galactic plane 
(their Figure 27), the Hercules Thick Disk 
cloud may be a debris stream consistent with
the disk formation scenario described by \cite{abd03}.  
While the Hercules Thick Disk Cloud is relatively nearby on the sky it
is not related to the more distant (10 - 20 kpc) Hercules-Aquila cloud of
\citep{bel07}. The northern
extent of the Hercules-Aquila cloud (Figure 2 in \citet{bel07}) is confined to
galactic latitudes less than $30\arcdeg$ and
even in those regions the bulk of the stars are much fainter than our
magnitude limit.  The contamination of our sample by Hercules-Aquila
cloud stars would be less than 5 objects per square degree (0.5 stars/bin)
in any case given our relatively bright magnitude limits.

Other explanations are still possible such as a triaxial thick disk 
and an overdensity due to an interaction with the disk bar.  
Analysis of our CCD photometry
and spectroscopy for fainter stars in Q1 and Q4 will be used to address
this question.

\acknowledgments

This work was supported by the National Science Foundation grants 
AST0507309 to Larsen and AST0507170 to Humphreys.
Larsen is pleased 
to thank Debora Katz and C. Elise Albert for useful conversations.

This research has made use of the MAPS Catalog of POSS I supported by the University of Minnesota. The APS databases can be accessed at http://aps.umn.edu/.

\clearpage

\end{document}